\documentclass[aps,prb,twocolumn,superscriptaddress,notitlepage,showpacs,amsmath,amstex,amssymb,citeautoscript,longbibliography,floatfix,a4paper,twocolumn]{revtex4-2}
\pdfoutput=1
\usepackage[english]{babel}
\usepackage{letltxmacro}
\usepackage{latexsym}
\LetLtxMacro{\ORIGselectlanguage}{\selectlanguage}
\makeatletter
\DeclareRobustCommand{\selectlanguage}[1]{%
  \@ifundefined{alias@\string#1}
    {\ORIGselectlanguage{#1}}
    {\begingroup\edef\x{\endgroup
       \noexpand\ORIGselectlanguage{\@nameuse{alias@#1}}}\x}%
}
\newcommand{\definelanguagealias}[2]{%
  \@namedef{alias@#1}{#2}%
}
\makeatother
\definelanguagealias{en}{english}
\definelanguagealias{English}{english}
\usepackage{graphicx}
\usepackage{amsmath}
\usepackage{amsfonts}
\usepackage{amssymb,bbding}
\usepackage{bm}
\usepackage{color}
\usepackage[percent]{overpic}
\usepackage{soul} 
\usepackage{amssymb}
\usepackage{wasysym}
\usepackage{dsfont}
\usepackage{float}
\usepackage{braket}
\usepackage{cancel}
\usepackage{esvect}
\usepackage{hyperref}
\hypersetup{
    bookmarks=false,         
    unicode=false,          
    pdftoolbar=false,        
    pdfmenubar=true,        
    pdffitwindow=false,     
    pdfstartview={FitH},    
    pdftitle={Entanglement transitions from Restricted Boltzmann Machines},    
    pdfauthor={R. Medina},     
    pdfsubject={},   
    pdfcreator={},   
    pdfproducer={}, 
    pdfkeywords={}, 
    pdfnewwindow=true,      
    colorlinks=true,       
    linkcolor=black,          
    citecolor=blue,        
    filecolor=magenta,      
    urlcolor=blue           
}

\newcommand{\be}{\begin{equation}}
\newcommand{\ee}{\end{equation}}
\newcommand{\bea}{\begin{eqnarray}}
\newcommand{\eea}{\end{eqnarray}}

\begin{document}
\title{Entanglement transitions from restricted Boltzmann machines} 
\author{Raimel Medina}
\affiliation{IST Austria, Am Campus 1, 3400 Klosterneuburg, Austria}
\author{Romain Vasseur}
\affiliation{Department of Physics, University of Massachusetts, Amherst, MA 01003, USA}
\author{Maksym Serbyn}
\affiliation{IST Austria, Am Campus 1, 3400 Klosterneuburg, Austria}

\date{\today}
\begin{abstract}
The search for novel entangled phases of matter has lead to the recent discovery of a new class of ``entanglement transitions'', exemplified by random tensor networks and monitored quantum circuits. Most known examples can be understood as some classical ordering transitions in an underlying statistical mechanics model, where entanglement maps onto the free energy cost of inserting a domain wall. In this paper, we study the possibility of entanglement transitions driven by physics beyond such statistical mechanics mappings. Motivated by recent applications of neural network-inspired variational Ans\"atze, we investigate under what conditions on the variational parameters these Ans\"atze can capture an entanglement transition. We study the entanglement scaling of short-range restricted Boltzmann machine (RBM) quantum states with random phases. For uncorrelated random phases, we analytically demonstrate the absence of an entanglement transition and reveal subtle finite size effects in finite size numerical simulations. Introducing phases with correlations decaying as $1/r^\alpha$ in real space, we observe three regions with a different scaling of entanglement entropy depending on the exponent $\alpha$.  We study the nature of the transition between these regions, finding numerical evidence for critical behavior. Our work establishes the presence of long-range correlated phases in RBM-based wave functions as a required ingredient for entanglement transitions. 
\end{abstract}
\maketitle
\section{Introduction}
The past decade has seen growing interest in identifying new phases of entangled, out-of-equilibrium quantum matter. A sufficiently generic, isolated quantum system is expected to approach thermal equilibrium under its unitary dynamics, and develop an extensive amount of entanglement entropy~\cite{Deutsch_PRA_1991, Srednicki_Entropy_and_area_1993, Srednicki_ETH_1994}. The eigenstates of such systems feature entanglement entropy scaling as a volume law. On the other hand, the presence of disorder or a quasiperiodic potential is capable of drastically changing the dynamical properties of the system, leading to many-body localization (MBL). MBL systems represent a different dynamical phase of matter, in particular characterized by area-law entangled eigenstates. The existence of distinct dynamical phases characterized by qualitatively different scaling of entanglement entropy naturally raises the question of the nature of the so-called MBL phase transition separating the ergodic and MBL phases. While the MBL transition received significant attention~\cite{Nandkishore_2015_review,colloquium_mbl_2019},  it was later realized to be a particular example of an \emph{entanglement transition}, which separates phases with different entanglement scalings. 

Beyond MBL,  examples of an entanglement transition are provided by measurement-induced entanglement transitions in random unitary circuits under random local measurements~\cite{Li_PRB_2018, Chan_PRB_2019, Skinner_PRX_2019, Bao_PRB_2020, Jian_PRB_2020, Gullans_PRX_2020, Zabalo_PRB_2020, Gullans_PRL_2020, Tang_PRR_2020, Choi_PRL_2020, Ippoliti_2021, PhysRevB.100.134306, Cao_2019, Nahum_2020_PRR, PhysRevX.11.011030, Lavasani_2021_Nature, sang2020measurement, li2020conformal, PhysRevB.102.014315, PhysRevLett.126.170602, Fuji_2020, PhysRevResearch.2.043072, Szyniszewski_2020, vijay2020measurementdriven, PhysRevB.103.104306, Fidkowski_2021, PRXQuantum.2.010352,lu2021entanglement,buchhold2021effective,bentsen2021measurementinduced,PhysRevLett.126.170503}, random tensor networks~\cite{Hayden:2016aa,Vasseur_2019_ET_RTN,2020arXiv201204666J,Lopez_Piqueres_2020,PRXQuantum.2.010352}, and Rokshar-Kivelson~\cite{RK_PRL_1988} (RK) inspired wave functions~\cite{Fradkin_2015_RK_and_EntanglementT}. 
While all these phase transitions feature different tuning parameters and phenomenology, they often can be mapped  onto classical ordering (spontaneous symmetry-breaking) phase transitions. For instance, the measurement-induced transition, resulting from a competition between local random unitary dynamics and the rate of local projective measurements at random points in space and time, was shown~\cite{Jian_PRB_2020, Bao_PRB_2020} to be related to a classical ordering transition in a replicated two-dimensional statistical model. Changing the measurement rate tunes the effective temperature of the classical model, driving it from an ordered phase at small measurement rate~(corresponding to volume-law entanglement) to a disordered phase at large measurement frequency~(area-law). Similar entanglement transitions and mappings also were reported in random tensor networks~\cite{Hayden:2016aa,Vasseur_2019_ET_RTN}.

 A different approach to entanglement transitions was considered in~\cite{Fradkin_2015_RK_and_EntanglementT}. There, the authors considered a variational Ansatz state inspired by the Rokhsar-Kivelson wave function~\cite{RK_PRL_1988}, in which the weights for the configurations are chosen from the Gibbs weights of a classical spin-glass model. Since states with positive weights can only sustain an area-law worth of entanglement~\cite{Grover_2015}, an additional random sign structure was also considered. The entanglement transition in this scenario was proven to be related to the geometric localization of the wave function due to the classical spin glass transition. 
 
Thus, on the one hand, both random tensor networks and measurement-induced entanglement transitions can be understood as an ordering transition in a two-dimensional statistical mechanical model, with the transition being induced by changing the tensor network bond dimension or the measurement rate. On the other hand, the transition observed in the RK-inspired wave function is closely related to a spin glass transition in an all-to-all spin-glass model. This motivates the question if entanglement transitions can be driven by physics beyond classical ordering transitions. The MBL transition as mentioned above could be one example of this. However, investigating the MBL transition is particularly challenging, see~\cite{colloquium_mbl_2019} and references therein for recent progress. 

In this work, we focus on the variational wave function approach. Specifically, in this paper, we consider a neural-network inspired variational Ansatz for the quantum wave function~\cite{Carleo_2017}. Machine learning techniques have been proved fruitful to the field of many-body quantum physics~\cite{Carleo_2017, Gao_NatureC_2017, arsenault2015machine, machine_learning_phases_of_matter, Wang_PRB_2016, Broecker_SR_2017, Kelvin_PRX_2017, Hibat_PRR_2020, Vieijra_2020,Levine_2019_QuantumEnt_and_DeepL_arch, Jia2019EntanglementAL, morawetz2020u1}. Specifically, exciting progress has been made in identifying quantum phases and transitions among them, either symmetry-broken phases~\cite{machine_learning_phases_of_matter, van_Nieu_NatureP_2017, Wang_PRB_2016, Broecker_SR_2017, Kelvin_PRX_2017} or topological phases~\cite{Deng_PRB_2017}, establishing connections to renormalization group techniques~\cite{beny2013deep, mehta2014exact} and perturbation theory~\cite{Borin}. In addition, machine learning ideas have also been used in measuring quantum entanglement and wave function tomography~\cite{tubman2016measuring, Palmieri_NPJ, morawetz2020u1}. We focus on a particular neural network variational state, referred to as restricted Boltzmann machines (RBM) quantum states which have been proven~\cite{DasSarma_2017, Jia2019EntanglementAL} to capture both an area-law as well as volume-law worth of entanglement depending on the locality of the neural network. In addition, recent work~\cite{Vieijra_2020} showed that RBM states can also be used to parametrize quantum wave functions with non-Abelian symmetries. Having in mind the representability power of neural network Ans\"atze, and in particular, of the simple RBM Ansatz, we ask under what conditions it is capable of capturing an entanglement transition while continuously varying some parameters used to define the state.

Since our motivation lies in entanglement transitions not driven by classical phase transitions, we consider RBM quantum states where the real part of the wave function is a local function of the spin degrees of freedom. Since no classical phase transition can exist in local one-dimensional statistical models, in this way we exclude any possible entanglement phase transition driven by a classical phase transition in the wave function weights. Our main results and structure of this paper are as follows: After introducing short-range RBM quantum states and discussing their entanglement capabilities in  Sec.~\ref{sec:Section2}, we proceed with the study of short-range RMB states with uncorrelated random phases in Sec.~\ref{sec:Section3}. There we demonstrate the volume-law entanglement scaling for all values of the variational parameter $\beta$ that plays a role of inverse effective temperature, indicating the absence of an entanglement transition. Next, in Sec.~\ref{sec:Section4}  we study the phase diagram for short-range RBM quantum states but with power-law correlated phases as a function of inverse temperature $\beta$ and power-law exponent $\alpha$.  We numerically show the existence of three different scaling regimes of the entanglement entropy and discuss the existence of a possible entanglement transition in the two dimensional phase diagram characterized by $\alpha$ and~$\beta$.  We conclude in Sec.~\ref{sec:Section5} with brief discussions of our results and comments on directions for future work.

\section{Restricted Boltzmann Machines}\label{sec:Section2}
In this section, we introduce restricted Boltzmann machines in the context of their usage as a variational state for many-body quantum physics. Moreover, we discuss the entanglement properties of the resulting wave function which depends on the locality of the RBM~Ansatz.

\subsection{RMB as a wave function Ansatz}
A restriction from a Boltzmann machine~\cite{Ackley1985-ACKALA-2}, the so-called restricted Boltzmann machine (RBM) is one of the simplest cases of an artificial neural network that can learn a distribution over the set of their input. In the context of quantum many-body physics~\cite{Carleo_2017}, RBMs are often used to approximate the wave function that is interpreted as a (complex) marginal probability distribution. Assuming that the wave function is defined on the many-body Hilbert space of spin-1/2 degrees of freedom, the spin configurations $\sigma\in \{-1,1\}$ represent an input referred to as ``physical spins''.

\begin{figure}[tb]
\begin{center}
\includegraphics[width=0.95\columnwidth]{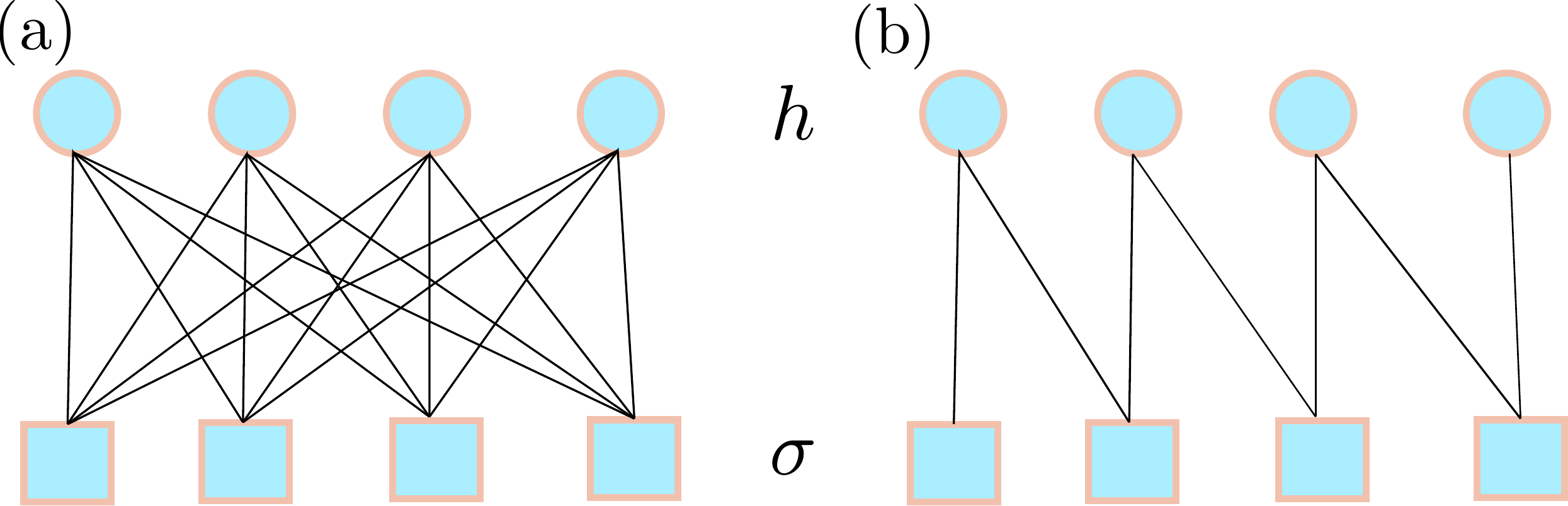}\\
\caption{\label{Fig:cartoon}
{\bf{General and local RBM.}} (a) Most generic connectivity of RBM with a dense interaction $W_{ij}$ matrix. (b) Example of ``local'' RBM network. Only nearest neighbour interactions between physical and hidden spins are considered.}
\end{center}
\end{figure}

The RBM is defined on a bipartite graph, whose vertices are grouped into two classes: the hidden spins and the physical (visible) spins, see Fig.~\ref{Fig:cartoon}.
Suppose there are $N$ visible spins and $M$ hidden spins, and we
associate the $N$ variables $\sigma\in\{-1,1\}$ and $M$ variables $h\in\{-1,1\}$ to the physical and hidden spins respectively. The spins in the hidden layer are connected to those in the physical layer, but there is no connection among spins in the same layer (see Fig.~\ref{Fig:cartoon} for an illustration). The RBM representation of the amplitudes of a quantum many-body wave function is obtained by tracing out the hidden spins:
\begin{equation}\label{Eq:amplitudes_RBM}
\Psi_{\Omega}(\sigma) = \sum_{h_j = \pm 1} e^{\sum^{N}_{i=1} a_i \sigma_i + \sum^M_{j=1} b_j h_j + \sum_{ij}\sigma_i W_{ij} h_j },
\end{equation} 
and is fully specified by a set of variational parameters denoted as $\Omega = \{W, a, b\}$ that can assume complex values. Here non-zero interactions between hidden and physical spins $W_{ij}$ induce correlations between physical spins, whereas constants $a_i$ renormalize the wave function amplitudes depending on the state of individual spins and do not induce correlations. The presence of either non-zero $a_i$ or $b_i$ leads to a breaking of the ${\mathbb Z}_2$ symmetry associated with flipping all physical spins. The summation over hidden spins in Eq.~\eqref{Eq:amplitudes_RBM} can be performed analytically, resulting in 
\begin{equation}\label{Eq:Psi-RBM}
\Psi_{\Omega}(\sigma)=e^{\sum_{i=1}^{N}a_{i}\sigma_{i}}\prod_{j=1}^{M}2 \cosh \left(b_{j}+\sum_{i=1}^N W_{ij}\sigma_{i} \right).
\end{equation}
Since the quantum many-body wave function typically contains complex parameters, we rewrite each term in the sum in its polar form $\Psi_{\Omega}(\sigma)=e^{i\phi_\sigma} e^{-E_{\Omega}(\sigma)/2}$, where both $\phi_\sigma$  and $E_{\Omega}(\sigma)$ are real, with the latter playing an analogue of an energy in the partition function that determines the norm of the state. More specifically, we introduce the partition function  $Z=\left. \langle \psi_{\Omega} \right| \psi_{\Omega} \rangle =\sum_\sigma e^{-E_\Omega(\sigma)}$, thus corresponding to a statistical model characterized by the energy function $E_\Omega(\sigma)$. In the remaining of the paper, we will explore the structure of entanglement in this wave function, depending on the different choices of the variational parameters $\Omega$. 

\subsection{Entanglement of RBM states}
As supported by representability theorems~\cite{LeRoux_2008}, RBM can approximate any many-body quantum state. However, such approximation may require an exponential in system size number of hidden spins and parameters, thus rendering the representation impractical. The interest lies then in finding types of many-body quantum states that can be \emph{efficiently} described by RBM Ansatz with a number of hidden spins and parameters scaling \emph{polynomially} in the number of physical spins $N$. Throughout this work, we restrict to Ans\"atze that have the same number of physical and hidden spins, $N=M$.

Entanglement is a crucial feature of the quantum wave function. Hence, to understand the representability power of Ans\"atze that are polynomial in system size, a study of the entanglement structure of such quantum states is required. The entanglement properties of RBM were first studied in~\cite{DasSarma_2017}, where it was demonstrated that RBM states with short-range interactions between hidden and visible spins feature an
area law of entanglement in any dimension and for arbitrary network geometry. More precisely, it was proven (see Theorem 1 of \cite{DasSarma_2017}) that when wave function~(\ref{Eq:Psi-RBM}) has $W_{ij}=0$ if $|i-j|>\mathcal{R}$,  Renyi entropies with an index $n$ of a region $A$ are bounded as $S_{n}(A)\leq (2 \mathcal{R}  \log 2)  \partial A$ for all values of $n$. Here $\partial A$ denotes the surface area of subsystem $A$ and the Renyi entropy with index $n$ is defined as 
$$
S_n(\rho_A) = \frac{1}{1-n}\ln \mathop{\rm{Tr}} \rho_A^n,
$$ 
where $\rho_A$ is a density matrix of the region $A$, obtained by tracing out all complementary degrees of freedom. This bound on all Renyi entropies indicates interesting connections between short-range RBM states and quantum states represented by a matrix-product state~(MPS)~\cite{Werner_1992, Cirac_2007, SCHOLLWOCK201196}. In particular, it was shown that all 1D short-range RBM~\cite{DasSarma_2017, Cirac_RBMs_2018}, as well as sufficiently sparse~\cite{Chen_RBM_and_TNS_2018} RBM can be efficiently described in terms of MPS. It is interesting to note that the validity of the inverse statement is to the best of our knowledge unknown, that is, given a generic MPS with a small bond dimension it is unknown whether there exists a representation in terms of a short-range RBM. 

Moreover, Ref.~\cite{DasSarma_2017} demonstrated via an explicit example that long-range RBM states can exhibit volume-law entanglement. This further implies that the corresponding MPS representation of such a state would require an exponential scaling of the bond dimension with the system size. 

As we discussed above, RBM Ans\"atze are capable of representing quantum states with both area- and volume-law entanglement. This suggests the natural hypothesis that RBM may provide a potential Ansatz for a variational wave function that exhibits an entanglement transition. Inspired by this hypothesis, in the remainder of this work we explore entanglement structure of the RBM quantum states and its dependence on the variational parameters.

\section{Uncorrelated random phases}\label{sec:Section3}
\begin{figure*}[t]
	\begin{center}
		\includegraphics[width=1.98\columnwidth]{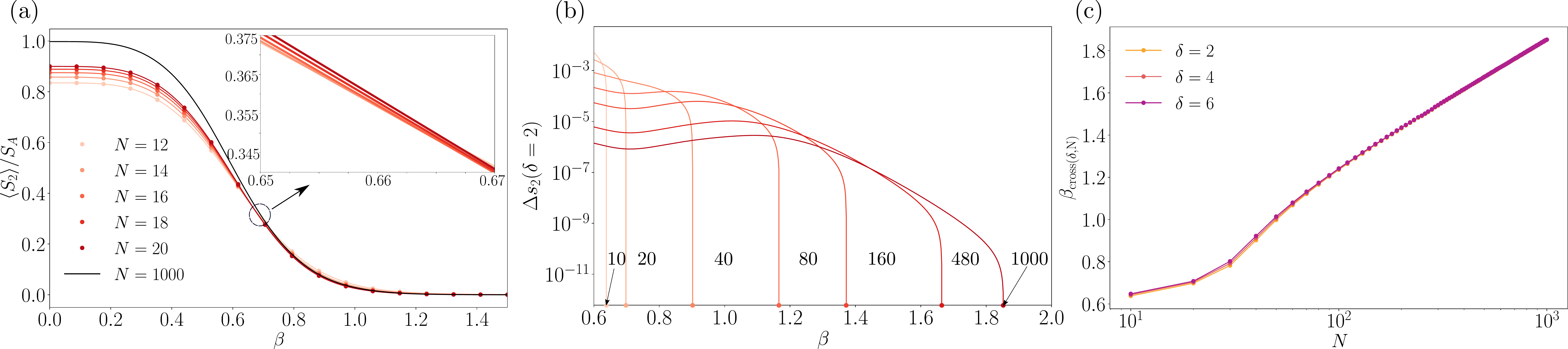}
		\caption{ \label{fig:constant_E_urandph} {\bf{Absence of entanglement transition in model with uncorrelated random phases.}} 
	(a) The apparent finite-size crossing in the plot $s_2 = \langle S_{2}\rangle/S_{A}$ vs $\beta$, for the RBM state with constant energy function and i.i.d random phases, {that could be attributed to a possible} entanglement transition. Points correspond to the quenched average computed numerically, and they nearly perfectly agree with the lower bound (lines) calculated analytically. {The inset shows the drift of the crossing point for analytic curves for different system sizes.}
(b) Difference of the 2nd-Renyi entropy density for system sizes $N$ and $N+2$, showing that the crossing points shift to higher values of $\beta$ with system size $N$, indicating the absence of a transition. (c) At large system sizes we observe a slow, logarithmic dependence of $\beta_{{\rm cross}}$ with system size, i.e., $\beta_{{\rm cross}}(\delta,N)\sim\ln N$ in the $N\gg1$ limit, indicating the absence of a transition. Here we considered system size spacings $\delta =2, 4$ and $6$.
	}
	\end{center}
\end{figure*}

In this section we motivate and introduce the first family of variational RBM Ans\"atze with random uncorrelated phases. After that we discuss results for the annealed average of the second Renyi entropy which shows the absence of an entanglement transition in this family of wave functions.

\subsection{RBM ansatz wave function}
In this work, we only consider ansatzes that do not allow for a description of the entanglement transition in terms of a classical phase transition in the wave function weights. Is it then natural to restrict to local RBM quantum states. However, such construction cannot encode volume law entanglement due to result of Ref.~\cite{DasSarma_2017} discussed above. Thus we introduce a minimal extension to the short-range RBM ansatz and include an additional contribution coming from independent and identically distributed (i.i.d) random phases. In this aspect, our construction is similar to that of Ref.~\cite{Fradkin_2015_RK_and_EntanglementT} that extended the variational wave function of the Rokshar-Kivelson type with random phases. Crucially, here we keep the interactions between spins strictly local.

Specifically, we consider the following ansatz state as our starting point:
\begin{align}
|\psi_\Omega(\beta)\rangle & =\frac{1}{\sqrt{Z}}\sum_{\sigma}e^{i\phi_{\sigma}}\Psi_\Omega(\sigma,\beta)|\sigma\rangle,
\label{eq:sRBM}
\end{align}
where each phase $\phi_{\sigma}$ is i.i.d and drawn from a uniform distribution in the $[-\pi, \pi]$ interval. This wave function is parametrized by the parameter $\beta$ that plays the role of effective inverse temperature in the RBM-type state $\Psi_\Omega(\sigma,\beta)$,
$$
\Psi_\Omega(\sigma,\beta)=\prod_{j}^{N}\cosh\big(\beta(\lambda+\sigma_{j}+\sigma_{j+1})\big).
$$ 
This state is a particular example of RBM wave function from Eq.~(\ref{Eq:Psi-RBM}), 
where the network geometry includes only nearest neighbor interactions between physical and hidden spins, see Fig.~\ref{Fig:cartoon}(b). We note that $\Psi_\Omega(\sigma,\beta)>0$ for all states $|\sigma\rangle$ in the computational basis, and hence it can be written as $\Psi_\Omega(\sigma,\beta) =e^{-E_{\Omega}(\beta)/2}$,
where 
\begin{equation}\label{eq:energy_srbm}
E_\Omega(\beta)=-2\sum_{j}\ln\cosh\big(\beta(\lambda+\sigma_{j}+\sigma_{j+1})\big),
\end{equation}
 can be interpreted as the energy of some underlying statistical mechanical model. In what follows we fix the value of lambda such that $\lambda=0.1$.

\subsection{Bounding Renyi entropies}
We are interested in studying average Renyi entropies, 
\begin{equation}
\langle S_{n}(\rho_{A})\rangle=\frac{1}{1-n}\big(\langle\ln{\rm Tr}\big(\bar{\rho}_{A}\big)^{n}\rangle-n\langle\ln Z\rangle\big),\label{eq:quenched_ent}
\end{equation}
where $\rho_A=\frac{1}{Z}\bar{\rho}_A$ represent the normalized quantum state of the subsystem $A$. The averaging is performed over the random phases $\phi_\sigma$ in Eq.~(\ref{eq:sRBM}). However, in general calculating $\langle\ln\bar{\rho}_{A}^{n}\rangle$ is analytically intractable and hence we resort to computing a lower bound. The lower bound, $\langle S_{n}(\rho_{A})\rangle
\geq\langle S_{n}(\rho_{A})\rangle_{{\rm ann}}$, uses the so-called annealed average of the same Renyi entropy
\begin{equation}
\langle S_{n}(\rho_{A})\rangle_{{\rm ann}}=\frac{1}{1-n}\big(\ln\langle{\rm Tr}\big(\bar{\rho}_{A}\big)^{n}\rangle-n\ln \langle Z\rangle\big),
\label{eq:anneal_ent}
\end{equation}
where the disorder-averaged density matrix and partition function enter the logarithm.

It is possible to show that for any state of the form of Eq.~\eqref{eq:sRBM} all annealed Renyi entropies can expressed via modified partition functions depending on the Renyi index $n$. In particular, focusing on the second  Renyi entropy we obtain 
\begin{equation}
\langle S_{2}(\rho_{A})\rangle_{{\rm ann}}=-\ln\left(\frac{{\rm Tr}_{A}Z^2_A+{\rm Tr}_{B}Z^2_B}{Z^{2}(\beta)}-Y(\beta)\right),\label{eq:lower_sRBM}
\end{equation}
where $Z_{A,B} = {\rm Tr}_{A,B} e^{- E_\Omega(\beta)}$ are the partition function on region $A$ and its complement $B$, and $Y= Z(2\beta)/Z^2(\beta)$ corresponds to the inverse participation ratio. Such representation of the second Renyi entropy opens the door to analytical calculations of Eq.~\eqref{eq:lower_sRBM}. The final results is then an  analytical function of the system size $N$ and the variational parameter $\beta$. This allows us to obtain the value of $\langle S_{2}(\beta)\rangle_{{\rm ann}}$ for an arbitrary system sizes $N$. We relegate the detailed derivation to Appendix~\ref{App:A} and instead discuss results below. 

\subsection{Absence of entanglement transition}
In Fig.~\ref{fig:constant_E_urandph}(a) we compare the numerical results for the average bipartite second Renyi entropy with the corresponding lower bound. We observe a good agreement between the average second Renyi entropy and its lower bound throughout the full range of considered values of $\beta$. At low $\beta$ the bipartite second Renyi entropy scales extensively with system size, $S_2 \sim N$ which further implies the same scaling for the entanglement entropy. At larger values of $\beta$ we observe an apparent crossing for curves corresponding to different system sizes. While this may suggest the existence of an entanglement transition tuned by $\beta$, below we will argue that this crossing originates from strong finite size effects. 

To explore the fate of the crossing between normalized second Renyi entropy curves as the number of spins is increased, we first compute the difference between the annealed second Renyi entropy for systems consisting of $N$ and $N-2$ spins. Figure~\ref{fig:constant_E_urandph}(b) shows that the crossing point shifts toward larger values of $\beta$ when the number of spins $N$ is increased, suggesting the absence of an entanglement transition in the present case. The absence of transition is further supported by Fig.~\ref{fig:constant_E_urandph}(c), that shows a logarithmic increase of the value of $\beta_\text{cross}^{(\delta,N)}$ defined as the point where crossing occurs between normalized second Renyi entropy for $N$ and $N+\delta$ spins. The logarithmic increase with $N$ holds for $\beta_\text{cross}^{(\delta,N)}$ irrespective of the separation $\delta$ between the system sizes considered, giving a strong indication that the crossing shifts to $\beta=\infty$ in the thermodynamic limit. 

Summarizing, the above results imply that short-range RBM quantum states with uncorrelated random phases, Eq.~\eqref{eq:sRBM}, always display a volume-law entanglement at arbitrary large but finite values of $\beta$. Therefore this ansatz does not have an entanglement transition as a function of $\beta$. While the result here was obtained for a simple class of RBM states with only nearest neighbor interactions, we also numerically confirmed it for longer-range local RBM states. Thus we believe that arbitrary local (finite-range) RBM states feature volume-law entanglement for any finite values of $\beta$, and thus do not have an entanglement transition. 

\section{Correlated random phases}\label{sec:Section4}

\begin{figure*}
\includegraphics[width=1.9\columnwidth]{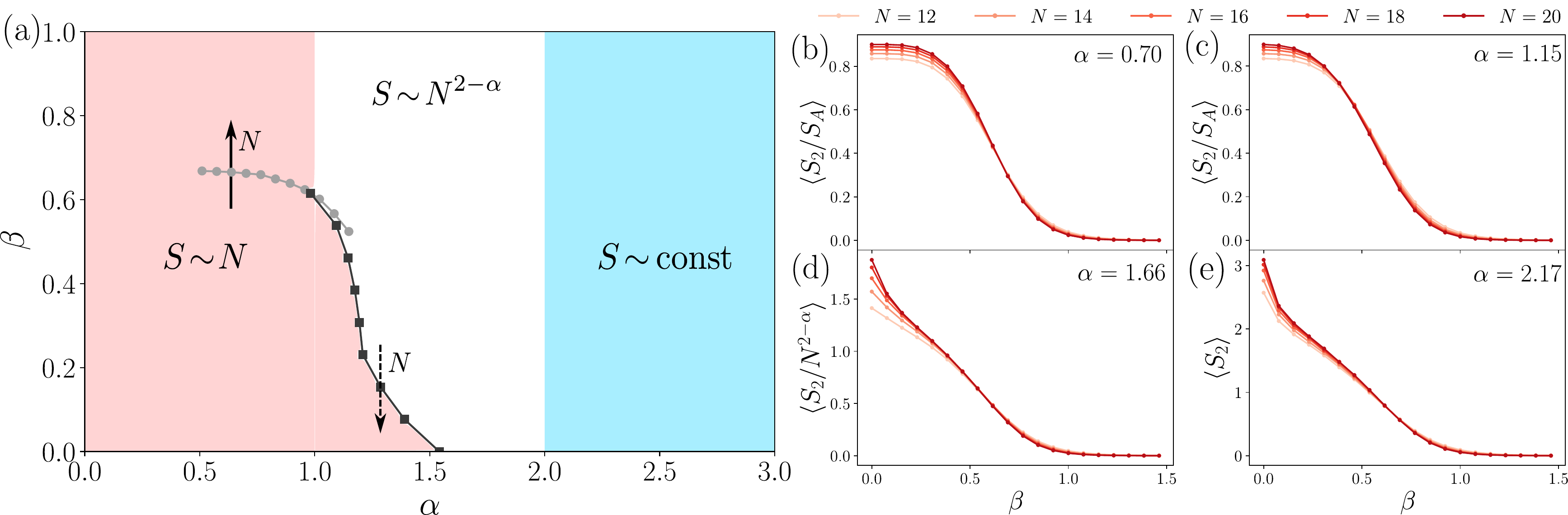}
\caption{{\bf{Phase diagram:}} (a) The phase diagram derived from the implications of the analytical bound, which implies $S\sim N^{2-\alpha}$ for $\alpha<2$  and $S\sim \text{const}$ for $\alpha>2$ for large values of $\beta$. For smaller $\beta$ we use the finite size scaling of the numerical data, which lead to the tentative critical line $\beta_c=\beta(\alpha)$ (grey and black).  Here the arrows refer to the different finite size behaviors of the tentative critical line $\beta_c$, and could be consistent with its disappearance in the thermodynamic limit.  The use of different colours in the critical line for $\alpha<1$ (grey) and $\alpha>1$ (black) is to signal that the former corresponds to a crossover, and hence is it a finite size effect.
(b-e) Illustration of finite size scaling of $\langle S_2 \rangle/S_A$ for different fixed values of $\beta$ and number of physical spins ranging between $N=12$ and $N=20$.  
(b-c) For values of $\alpha<1.5$ the presence of crossings is observed suggesting the existence of a critical line separating two regions with different scaling for the entanglement entropy. However, the stability of this observation in the thermodynamic limit cannot be fully established within the current reach of our numerics; see the main text for a further discussion.
(d) For $1.5 < \alpha < 2$ numerics suggest the entropy to scale sub-extensively with system size $N$ in agreement with the analytical bounds predicting the entanglement entropy to be, at large values of $\beta$, upper bounded by $N^{2-\alpha}$.
(e) For $\alpha>2$ numerical results show the entropy becomes independent of system size $N$ hence satisfying an area-law of entanglement. 
}
\label{fig:phase_diag}
\end{figure*}

As we discussed in Sec.~\ref{sec:Section3}, the presence of uncorrelated random phases does not allow for an entanglement transition in the local RBM variational Ansatz. Two natural directions towards an entanglement transition consist in relaxing either the locality of the energy function $E_{\Omega}(\beta)$ or introducing correlations between random phases in Ansatz~(\ref{eq:sRBM}). In order to avoid a spin-glass transition in the classical Hamiltonian, leading to the geometric localization of the wave function, we keep the local structure of RBM and relax the condition of uncorrelated random phases. In particular, below we consider a short-range RBM ansatz with power-law correlated random phases, controlled by a power-law exponent $\alpha$.
\subsection{RBM Ansatz wave function}
We introduce correlations between random phases using the RBM description of the wave function with complex parameters. Specifically, we consider the Ansatz Eq.~(\ref{Eq:Psi-RBM}) with $a_i=0$ and complex values of $W_{ij}$ written as $W_{ij} =  W^{(1)}_{ij}+ i W^{(2)}_{ij}$. The real part of the couplings is taken to be local with constant nearest neighbors interaction, that is, $W^{(1)}_{ij}$ is the same as in Eq.~\eqref{eq:sRBM}. The imaginary part of couplings,  $W^{(2)}_{ij}$, is a random number that is suppressed with distance between points $i$ and $j$, 
\begin{equation}\label{Eq:}
W_{ij}^{(2)} = \frac{w_{ij}}{|i-j|^{\alpha}}, \text{ with $w_{ij} \in [-\pi, \pi]$.}
\end{equation}
 Here the power-law exponent $\alpha>0$ controls the decay of correlations in real space. In addition, we consider a longitudinal magnetic field of the form $b_j=\lambda + i b_j^{(2)}$, with $b_j^{(2)} \in [-\pi, \pi] $. 
 
 Translating the above choice of complex $W_{ij}$ into the wave function and keeping only the local part of the energy function we obtain the following expression:
\begin{equation}\label{eq:cRBM_state}
|\psi_\Omega(\beta)\rangle=\frac{1}{\sqrt{Z}}\sum_{\sigma}e^{i\sum_{j=1}^N\arg(z_{j}(\beta))}\prod_{j=1}^N\cosh(\beta X^{(1)}_{j})
|\sigma\rangle.
\end{equation}
where $z_j$ depends on the  complex parameters of the RBM under consideration,
\begin{equation*}
z_j = \cosh (\beta X^{(1)}_{j}) \cos X^{(2)}_{j}+i\sinh (\beta X^{(1)}_{j}) \sin X^{(2)}_{j},
\end{equation*}
with $X^{(1,2)}_j = b_{j}^{(1,2)}+\sum_{i=1}^N\sigma_{i}W_{ij}^{(1,2)}$. From this representation we see that the presence of non-zero imaginary part of the RBM parameters translates into correlated phases entering in the expression for the RBM wave function, Eq.~(\ref{eq:cRBM_state}).

The presence of non-local random phases in this model makes it impossible to find an exact expression for the lower bound of $\langle S_2(\beta) \rangle$ as done in Section.~\ref{sec:Section3}. Hence, we resort to perturbative expansions in the the high-$\beta$ regime of the state in Eq.~\eqref{eq:cRBM_state}.

\subsection{High-$\beta$ limit}
In the limit of large $\beta$ we expand Eq.~(\ref{eq:cRBM_state}) to first non-trivial order in $\beta|X^{(1)}_{j}|$, which gives following expression
\begin{equation}
\label{eq:high_beta_state_approx}
|\psi_\Omega(\beta\gg1)\rangle\approx \frac{1}{\sqrt{Z(\beta)}}\sum_{\sigma}e^{i\phi_{\sigma}}e^{-E_\Omega(\beta)/2}|\sigma\rangle,
\end{equation}
where $\phi_{\sigma}= \sum_{j=1}^{N}\mathop{\rm sign}[X^{(1)}_{j}]X^{(2)}_{j}$ and $E_\Omega(\beta)$ is given by Eq.~\eqref{eq:energy_srbm}. 
We can think of the state Eq.~\eqref{eq:high_beta_state_approx} as the result of time evolving the state 
$
|\Phi\rangle=\frac{1}{\sqrt{Z}}\sum_{\sigma}e^{-E_\Omega(\beta)/2}|\sigma\rangle
$ that has no phases with a Hamiltonian $H_{Z}$ that is diagonal in the computational basis and has eigenvalues 
$\phi_{\sigma}$. Since the state $|\Phi\rangle$ features an area-law entanglement~\cite{DasSarma_2017}, the structure of entanglement of the state $|\psi_\Omega(\beta\gg1)\rangle$ is determined by the phases encoded in $H_Z$. Below we will consider the structure of $H_Z$ and use it to understand the entanglement of the resulting state.

The explicit form of $H_{Z}$  depends on the range $\mathcal{R}$ of the real part of RBM couplings $W^{(1)}_{ij}$. However, for any finite $\cal R$ it features long-range power-law interactions between spins determined by the parameter $\alpha$. As an example, it is possible to show that for the present ansatz 
\begin{equation}
H_Z =\sum_{ij}\frac{J_{ij}}{|i-j|^\alpha}  \sigma^z_i\sigma^z_j + \sum_i h_i \sigma^z_i,
\end{equation}
where $J_{ij} \propto w^2_{ij}$ and $h_i$ is a random number uniformly distributed in the $[1/2, 1/2+\pi]$ interval. In order to understand the entanglement created by $H_Z$, we use results of Ref.~\cite{Gorshkov_2017_Entanglement_area_laws}. This work obtained an upper bound on the rate of production of entanglement entropy, $\Gamma = dS_A/dt$. The entanglement rate scales as the area of the boundary's subregion when $\alpha>2$ ($\alpha>D+1$ in arbitrary dimensions). Since the initial state $|\Phi\rangle$ is an area-law state, the aforementioned result guarantees that the state $|\psi_\Omega(\beta\gg1)$ has an area-law entanglement if $\alpha>2$. In contrast, for $\alpha<2$ the results of Ref.~\cite{Gorshkov_2017_Entanglement_area_laws} imply that the entangling rate that scales as $\Gamma \sim N_A^{2-\alpha},$ where $N_A$ is the size of the subregion $A$ which here we always take as $N_A=N/2$ since we focus on the bipartite entanglement. Note that when $\alpha\in(1,2)$ we get that $2-\alpha<1$ and hence, the entangling rate scaling is sub-extensive in the system size $N$.

The application of the upper bound from Ref.~\cite{Gorshkov_2017_Entanglement_area_laws} results in the high-$\beta$ phase diagram boundaries shown at the top part of Fig.~\ref{fig:phase_diag}(a). For $\alpha>2$ the bound implies area-law scaling of entanglement. For the intermediate values of $1<\alpha<2$ we expect at most a subextensive amount of entanglement, $S\propto N_A^{2-\alpha}$. Finally, for the values of $\alpha<1$ the upper bound has volume-law entanglement and thus is not very useful. 

\subsection{Phase diagram}

Although the bound discussed above provides expectations for the phase diagram at high values of $\beta$, the expansion breaks down at small values of $\beta$. Therefore, we resort to numerical simulations to determine the phase boundaries at intermediate and small values of $\beta$. To be consistent with our previous approach, we focus on the second Renyi entropy and perform numerical simulations for systems with up to $N=20$ physical spins. The results of these simulations are summarized in Fig.~\ref{fig:phase_diag}.

To begin with, let us explore the region determined by $\alpha<1$, where according to summary shown in Fig.~\ref{fig:phase_diag}(a) we expect the volume law entanglement. In this region, the analytical bound does not provide any useful information. We compare the numerical results for average second Renyi entropy $\langle S_2(\alpha, \beta) \rangle$ with fixed $\alpha<1$ to the analytical result from Eq.~\eqref{eq:lower_sRBM} for the annealed average obtained under assumption of i.i.d random phases and the same non-random energy function. Remarkably, we observe a good agreement between both quantities at all values of $\beta \in [0, 1.5]$ considered in the numerical simulations, see an example in Fig.~\ref{fig:phase_diag}(b). In particular, we find that for every $\beta$ in the considered range the difference between the analytical expression and numerically averaged second Renyi entropy is on average $\sim 10^{-7}$ and assumes the maximal value of $\sim 10^{-4}$ in the worst case, achieved for small values of $\beta$. This result suggests that, in this regime, the average second Renyi entropy in the presence of long-range phase correlations nearly coincides to its annealed average value in the case of uncorrelated random phases. This case was considered in Section~\ref{sec:Section3}, where we demonstrated an absence of entanglement transition. 

\begin{figure}[t]
\includegraphics[width=\columnwidth]{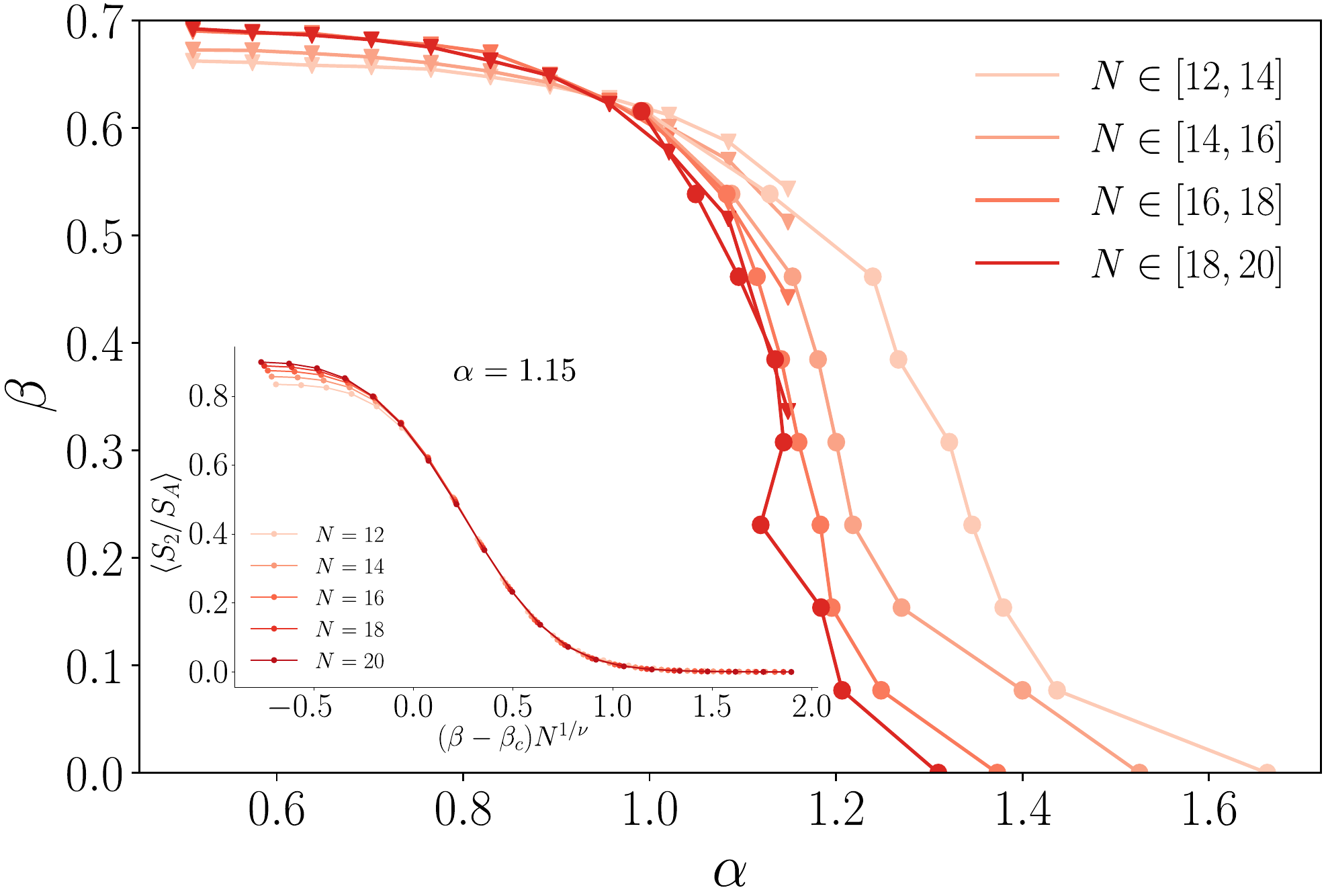}
\caption{{\bf Finite size scaling for the critical curve:} Clearly distinct behavior in the finite size scaling results for the critical $\beta(\alpha)$ ($\alpha(\beta)$) is observed for regions $\alpha \lessgtr 1$. For $\alpha<1$ with increasing system size the crossing point shifts towards larger $\beta$, which is in agreement with our previous expectations. Moreover, for $\alpha>1$ with increasing $N$ the crossing point shifts towards smaller $\alpha$. Circles (triangles) correspond to cuts at constant $\beta$ ($\alpha$). \emph{Inset}: Example of scaling collapse of numerical data at fixed $\alpha=1.15$. We find the critical exponets to be $(\beta_c, \nu) =(0.42, 0.5)$. Scaling collapses of data were performed using the \texttt{pyfssa}~\cite{pyfssa, autoscalepy} package.}
\label{fig:betac_vs_alpha_Rhalf}
\end{figure}

Therefore, although the finite size scaling reveals a crossing in  Fig.~\ref{fig:phase_diag}(b), we expect this crossing to shift to larger values of $\beta$ with increasing system size. 
This is further supported by Fig.~\ref{fig:betac_vs_alpha_Rhalf}, that shows the increase of the point $\beta_{\rm cross}$,  defined at the point where curves corresponding to different system sizes cross, for $\alpha<1$. This result is along the line with our expectations of a volume law entanglement at all finite $\beta$.  However, the available range of $N$ is insufficient to check if $\beta_\text{c}$ grows logarithmically with $N$ as we demonstrated in Sec.~\ref{sec:Section3}.

Next, we continue with the region $1 < \alpha <2$. Here the bound suggests the existence of a sub-extensive scaling of the entropy at large $\beta$, as indicated at the top of Fig.~\ref{fig:phase_diag}(a). Numerical simulations in Fig.~\ref{fig:phase_diag}(c) show the presence of crossings for values of $\alpha$ that are sufficiently close to one, for example $\alpha=1.15$. For larger values of $\alpha$, for instance $\alpha = 1.66$ no crossing is observed and instead we find a good agreement with a sub-extensive $N^{2-\alpha}$ scaling of the entropy shown in Fig.~\ref{fig:phase_diag}(d). 
This result suggests the existence of a critical region separating the volume-law and sub-extensive scaling regimes, as is illustrated by the black line in Fig.~\ref{fig:phase_diag}(a) that shows the position of the crossing. The inset of Fig.~\ref{fig:betac_vs_alpha_Rhalf} shows the scaling collapse for $\alpha=1.15$ with critical exponents $(\beta_c, \nu) = (0.42, 0.5)$. 

We observe the crossing and performed scaling collapse for the region of $1<\alpha<1.53$ that is suggestive of entanglement transition. However, as observed in Fig.~\ref{fig:betac_vs_alpha_Rhalf},  finite size scaling of data shows that in the region $\alpha>1$ the critical line shifts toward smaller $\alpha$ when considering larger system sizes. Such finite size drift may again suggest this region might be associate with a crossover rather than an actual entanglement transition, but due to our limited system size points, we cannot rule out any of the two possibilities.  

Finally, for  $\alpha>2$, numerical simulations in Fig.~\ref{fig:phase_diag}(e) show the entanglement entropy to be independent of system size for all finite values of $\beta$, signaling the existence of area-law scaling for the entanglement entropy. This is reflected in Fig.~\ref{fig:phase_diag}(a) by the shaded region for $\alpha>2$. 

In summary, the phase diagram proposed in Fig.~\ref{fig:phase_diag}(a) agrees with our numerical findings for the existence of three different scaling regimes for the entanglement entropy. The possible entanglement transition tuned by parameter $\beta$ could be realized in this phase diagram for an intermediate range of $\alpha$, $1 <  \alpha \lesssim 1.53$. However, the example of strong finite-size effects and eventual absence of transition for $\alpha < 1$ suggests that one has to interpret these findings with care, and additional studies maybe needed to rule out the possibility of this being a crossover rather than a true entanglement transition. 

\section{Discussions}\label{sec:Section5}
Motivated by recent studies of entanglement transitions, we considered a neural-network inspired variational Ansatz for the quantum wave function. We used RBM quantum states, which have been proven capable of representing ground states of many-body Hamiltonians with polynomial-size gap and quantum states generated by any polynomial-size quantum circuits~\cite{Gao_NatureC_2017,  Huang_2017}. We studied whether such variational states can capture an entanglement transition while continuously varying some parameters used to define the underlying neural network. 

First, we considered a short-range RBM state with independent random phases parametrized by $\beta$  that plays the role of an inverse temperature. For this Ansatz we analytically calculated a lower bound on the second Renyi entropy.  We found numerically that the averaged second Renyi entropy is very close to the analytic lower bound for all numerically accessible number of spins (up to $N=20$). Although both the analytical lower bound and the numerically averaged Renyi entropy display finite size crossings, we demonstrate that they are associated with a crossover rather than an actual entanglement transition. Using the analytical lower bound we observe that the crossing point $\beta_{\rm{cross}}$ scales very slowly, as $\ln N$ for large $N$, a scaling that would be extremely hard to recognize from numerical simulations due to the limited range of accessible number of spins. We conclude that the short-range RBM Ansatz with random phases features a volume law entanglement at all finite values of $\beta$, thus ruling out the possibility of an entanglement transition.

As a more general example, we relaxed the assumption of uncorrelated random phases, and considered random power-law correlated phases controlled by the exponent $\alpha$. The presence of long-range correlated phases did not allow to apply the analytic approach and we instead resorted to numerical simulations of the second Renyi entropy. Using known bounds on the entangling rate for systems with power-law interactions, we derived a hypothetical phase diagram with three regions of different scaling of entanglement entropy for large values of $\beta$. For small and intermediate values of $\beta$, our numerical simulations suggest the existence of a critical curve in the $(\beta, \alpha)$ plane. However, the limited range of number of spins we can access does not allow us to exclude the presence of a crossover rather than an actual phase transition. 

As a future research direction, it would be interesting to go beyond RBM states. In particular, a recent work~\cite{Levine_2019_QuantumEnt_and_DeepL_arch} suggested  that contemporary deep learning architectures, in the form of deep convolutional and recurrent networks, can also efficiently represent highly entangled quantum systems. Moreover, it was shown that such architectures can support volume-law entanglement scaling, polynomially more efficiently $\mathcal{O}(N)$ than presently employed RBM, $\mathcal{O}(N^2)$. Hence, it would then be interesting to carry out the same study we pursued in this work but using these state-of-the-art neural-network base wave functions.

\section*{Acknowledgments}
We would like to thank to S. De Nicola and P. Brighi for fruitful discussions and valuable feedback on the manuscript. R.M. and M.S. acknowledge support by the European Research Council (ERC) under the European Union's Horizon 2020 research and innovation program (Grant Agreement No.~850899). R.V. acknowledges support from the US Department of Energy, Office of Science, Basic Energy Sciences, under Early Career Award No. DE-SC0019168, and the Alfred P. Sloan Foundation through a Sloan Research Fellowship.

\appendix
\section{Analytic calculation of the annealed second Renyi entropy \label{App:A}}
The goal of this section is to explicitly derive the expression~\eqref{eq:lower_sRBM} for the annealed average of the second Renyi entropy. As we already stated, this results holds for states of the form:
$$
|\psi(\beta) \rangle = \frac{1}{\sqrt{Z}}\sum_{\sigma} e^{-i \phi_\sigma} e^{-E(\beta)/2} |\sigma \rangle,
$$
where $E(\beta)$ is a non-random function, and $\phi_\sigma$ is i.i.d uniformly distributed in the $[-\pi, \pi]$ interval. 

We are then interested in computing the following quantity 
\begin{equation}\nonumber
S_{2}^{A}(\rho_{A})=-\ln\langle{\rm Tr}\rho_{A}^{2}\rangle.
\end{equation}
Given a bipartition of the system $(A,B)$, the reduced density matrix is given by the following expression
$$\rho_{A}=\frac{1}{Z}\sum_{a,a'}c_{a,a'}|\sigma_{a}\rangle\langle\sigma_{a'}|,$$ and consequently
\begin{equation}\label{eq:S2append}
	\begin{split}
{\rm Tr}\rho_{A}^{2} &=\frac{1}{Z^{2}}\sum_{a,a'}c_{a,a'}c_{a',a} \\
&=\frac{1}{Z^{2}}\sum_{a,a'}\sum_{b,c}r_{a,b}r_{a',b}^{*}r_{a,c}^{*}r_{a',c}e^{-\frac{\beta}{2}\big(E_{a,b}+E_{a',b}+E_{a,c}+E_{a',c}\big)}
	\end{split}
\end{equation}
where $r_{a,b} = e^{-i \phi_{a,b}}$. In order to average over disorder realizations we need to compute $\langle r_{a,b}r_{a',b}^{*}r_{a,c}^{*}r_{a',c}\rangle$. For this, note that since $r_{a,b}$ is a phase the only way to get a non-zero contribution is by cancelling out the phases. Hence, it is quite easy to check that
$$
\langle r_{a,b}r_{a',b}^{*}r_{a,c}^{*}r_{a',c}\rangle = \delta_{a,a'}+\delta_{b,c}-\delta_{a,a'}\delta_{b,c}.
$$
Finally, substituting the above expression into Eq.~\eqref{eq:S2append} leads to the desired result
\begin{equation*}
\langle{\rm Tr}\rho_{A}^{2}\rangle	=\frac{{\rm Tr}_{A}\big({\rm Tr}_{B}e^{-\beta H}\big)^{2}+{\rm Tr}_{B}\big({\rm Tr}_{A}e^{-\beta H}\big)^{2}}{Z^{2}(\beta)}-Y_{2}(\beta).
\end{equation*}

\end{document}